\documentclass[aps,pra,twocolumn,preprintnumbers,floatfix]{revtex4-1}
\usepackage{graphicx}
\usepackage{dcolumn}
\usepackage{bm}
\usepackage[normalem]{ulem} 
\usepackage{latexsym,epsfig}
\usepackage{graphicx}
\usepackage{verbatim}
\usepackage{comment}
\usepackage{amsmath}
\usepackage{amssymb}
\usepackage{stmaryrd}
\usepackage{color}
\usepackage{epstopdf}
\usepackage{grffile}
\usepackage{float}
\usepackage{soul}
\usepackage{ulem}
\DeclareGraphicsExtensions{.eps}

\newcommand{\beq}{\begin{equation}}
\newcommand{\eeq}{\end{equation}}
\newcommand{\bea}{\begin{eqnarray}}
\newcommand{\eea}{\end{eqnarray}}
\newcommand{\ben}{\begin{eqnarray*}}
\newcommand{\een}{\end{eqnarray*}}
\newcommand{\bfig}{\begin{figure}}
\newcommand{\efig}{\end{figure}}

\usepackage{hyperref}
\hypersetup{
    colorlinks=true,      
    urlcolor=blue,
    citecolor=blue,
    linkcolor=blue
}

\begin{document}
\title{Flux induced re-entrant dynamics in the quantum walk of interacting bosons }
\author{Mrinal Kanti Giri$^{1}$, Biswajit Paul$^{2,3}$, and Tapan Mishra$^{2,3}$}
\email{mishratapan@gmail.com}
\affiliation{$^1$Centre for Quantum Engineering, Research and Education, TCG CREST, Salt Lake, Kolkata 700091, India\\
$^2$School of Physical Sciences, National Institute of Science Education and Research, Jatni 752050, India\\
$^3$Homi Bhabha National Institute, Training School Complex, Anushaktinagar, Mumbai 400094, India}

\date{\today}

\begin{abstract}
We study the QW of two interacting bosons on a two-leg ladder lattice in the presence of an artificial magnetic field. By considering an uniform flux piercing through the ladder, we show that in the limit of strong onsite repulsion and dominant rung-hopping, an initially slow dynamics becomes fast, then slow and fast again with increase in the flux strength indicating a re-entrant dynamics. This unusual behaviour is found to be associated with the formation, breaking and reformation of a bound pair state along the rung of the ladder. In addition to this we also find a re-entrant behaviour in the chiral dynamics where the chirality in the system first increases and then decreases with increase in interaction. We establish this unusual re-entrant behaviour in the dynamics by analysing the radial velocity, spreading of correlation, center-of-mass shearing and energy band diagrams. 
\end{abstract}

\maketitle

\section{Introduction} The study of dynamical evolution of a quantum state following a sudden quench of the system parameter is a topic of paramount interest in recent years\cite{Bloch2008, Moeckel_2010, Polkovnikov2011}. The quantum walk (QW) ~\cite{Aharonov}, which is an unitary time evolution of a few particle quantum state provides a bottom-up approach to understand such dynamics~\cite{wang2013physical, ctqw_review1, ctqw_review2,Kempe2003}. Due to their versatility, the QW of a particles on a lattice has found its application in a wide range of systems ranging from fundamental physics, quantum technologies~\cite{Ambainis2003, Childs2004, child2009, Childs2013, Chakraborty2016, Ryan_QW, Lovett2010} leading to their experimental observation in various lattice systems~\cite{Schmitz2009,Zahringer2010,Karski2009,Peruzzo2010,Greiner_walk,Yan2019,Bromberg2009}. Apart from the non-interacting particles, the QW of more than one particles have been investigated to understand the effect of strong correlation and quantum statistics on the dynamics. Recent studies have revealed various interesting scenarios from the QW of interacting particles such as the Hanburry-Brown and Twiss type interference, bunching and anti-bunching of particles\cite{Greiner_walk,lahini2012qw, wen2021}, spin-charge deconfinement~\cite{Bloch2020,SantosSSH2018}, dynamics of magnon bound states~\cite{bloch_magnon_expt}, many-body localization~\cite{Zakrzewski2017,PhysRevA.94.023601,Crespi2013qw,Luis-Arya}, pairing due to competing interactions~\cite{giri2022nontrivial,mondalwalk,maria_doublon}, chiral dynamics~\cite{Tai_2017, maria2022} and topological properties~\cite{PhysRevA.95.013619,Demler2010topo,Xie2020,Obuse2013topo,LewensteinMCD, Ramasesh2017}, Bloch oscillation~\cite{FlachPRA2010,Zakrzewski2017,Greiner_walk,wen2021, Longhi_BO_2012,Johnson2014} etc.

Primarily, the dynamics of a quantum state is defined by the spreading of the wavepacket and the correlation function in the QW. While the single particle QW exhibits a faster spreading of the wavepacket in an uniform lattice, addition of perturbation such as disorder, topological effects or geometric frustration leads to the slower spreading~\cite{rakovszky2015localization,chandrashekar2011disordered,schreiber2011decoherence,Xie2020,PhysRevLett.119.130501,Gedik2015,Razzoli2020,Sajid2019,Armando2018,Zakrzewski2017,Gadway2017direct,Gadway2018a,Gadway2018b}. On the other hand, in the QW of two identical particles, interaction and the choice of initial conditions play crucial role which often results in interesting phenomena. The simplest example is the QW with an initial state of two bosons located at the same (nearest neighbour) site in a one dimensional lattice exhibits a fast to slow dynamics with an increase in the onsite (nearest neighbour) interaction due to the formation of the onsite (nearest neighbour) bound state. When the two particles are initially at the nearest neighbour sites, then the spreading does not slow down as a function of onsite interaction due to fermionization~\cite{Greiner_walk,lahini2012qw,Dias2016,Chaohong_etc}. 
However, the QW of identical bosons residing on the rung of a two-leg ladder exhibits a fast to slow spreading as a function of onsite interaction due to the rung-pair formation or rung-localization~\cite{Fan2020,Ye2019}.

\begin{figure}[t]
    \centering
    \includegraphics[width=0.75\columnwidth]{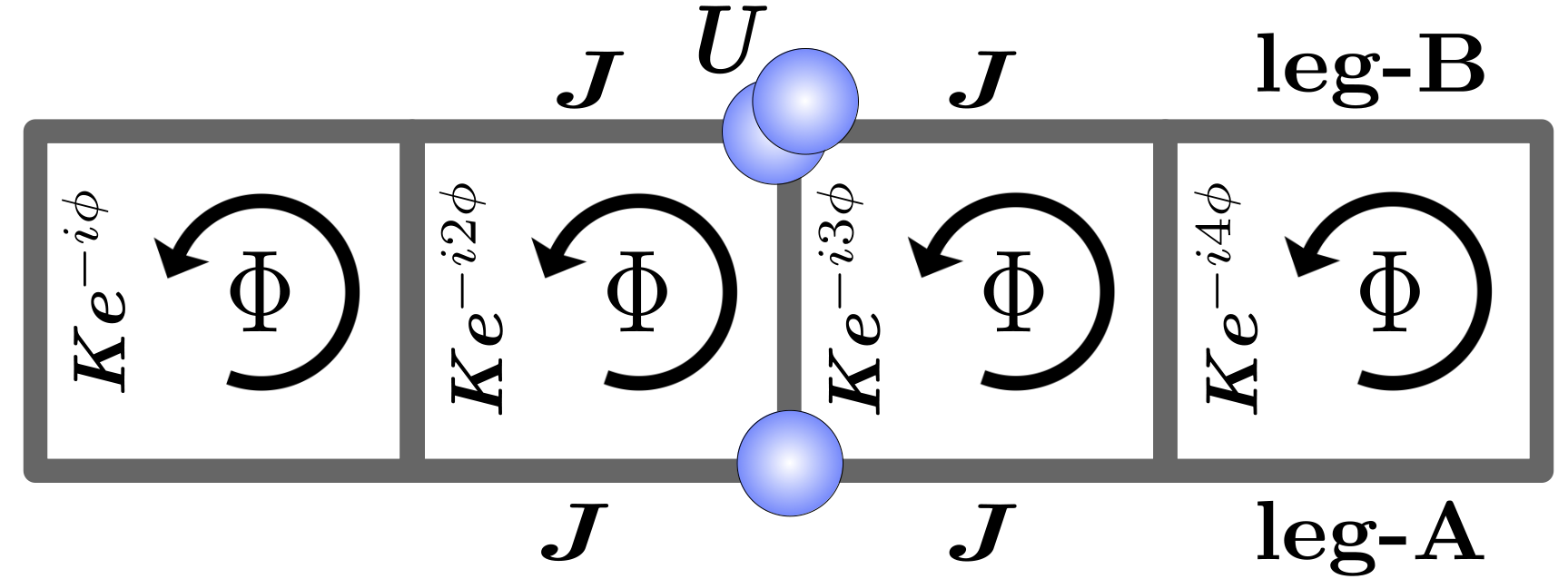}
    \caption{Figure depicting a two-leg BH ladder in the presence of uniform flux. $J$ and $K$ denote the intra and inter-leg hopping strengths and $U$ is the onsite two-body interaction. $\Phi$ is the flux piercing through each plaquette. }
        \label{fig:1}
\end{figure}
In all these cases, the QW exhibit an unidirectional or monotonous dynamics where an initially fast spreading becomes slow. However, in this paper we show that strongly interacting bosons on a two-leg ladder subjected to an artificial gauge field -a system known as the flux ladder~\cite{dhar2012, dhar2013, lehurflux2013,Tokuno_2014,lehurflux2015,Mishra2016,mishraflux2018, piraud2015vortex,greschner2015,greschner2016,Rashi_2017, Giamarchi2023,PhysRevA.106.063320,Sebastian2015,oktel2015} can exhibit a drastic deviation from this monotonous dynamics. 
By considering an initial state with two bosons residing on the central rung of the ladder, 
we show that in the regime of dominant rung hopping and strong onsite repulsion, an initial slow dynamics becomes fast, then slow and fast again as a function of flux piercing through the ladder. This indicates an interesting re-entrant behaviour in the dynamics which resembles a situation where an initial state transforms to another state which is microscopically similar or identical to the initial state. 
Furthermore, due to the presence of flux, we also find a re-entrant behaviour in the chiral dynamics as a function of interaction for fixed flux strengths. In the following, we elaborate on these findings in detail.

\section{Model}

The model describing the system of interacting bosons in a two-leg ladder subjected to uniform flux (Fig.~\ref{fig:1}) is given by the  Hamiltonian;
\begin{align}
    H &= \frac{U}{2}\sum_{l,\sigma} {n}_{l,\sigma}({n}_{l,\sigma}-1) -J\sum_{l,\sigma} \;({{b}}_{l, \sigma}^{\dagger} {b}_{l+1, \sigma} +h.c) \nonumber \\
    &-\;K\sum_{l}\; ({{b}}_{l,A}^{\dagger} {b}_{l,B}e^{-i l\phi}  + H.c.),
\label{eqn:ham1}
\end{align}
where, $b_{l,\sigma}$ ($b_{l, \sigma}^{\dagger}$) are bosonic annihilattion (creation) operators on the rung $l$ with the leg indices $\sigma\in A,B$ and ${n}_{l,\sigma}$ is the corresponding number operator. $J$ and $K$ denote the amplitudes of the intra-leg and inter-leg hopping strengths respectively. Due to the presence of the gauge field, the inter-leg hopping term is complex in nature which acquires a phase $\phi= \frac{e}{\hbar} \int_{r_{i}}^{r_{f}} d\mathbf{\vec{r}.A(\vec{r})}$, where $\mathbf{A(\vec{r})}$ is the magnetic vector potential.
We perform our calculations in the Landau gauge $\mathbf{A(\vec{r})} = Bx\hat{y}$ with the phase $\phi = \pi\Phi/\Phi_{0}$, where $\Phi$ is 
the magnetic flux through each plaquette and $\Phi_{0} = h/e$ is the magnetic flux quantum. 
\begin{figure}[t]
    \centering
    \includegraphics[width=1\columnwidth,height=0.15\textheight]{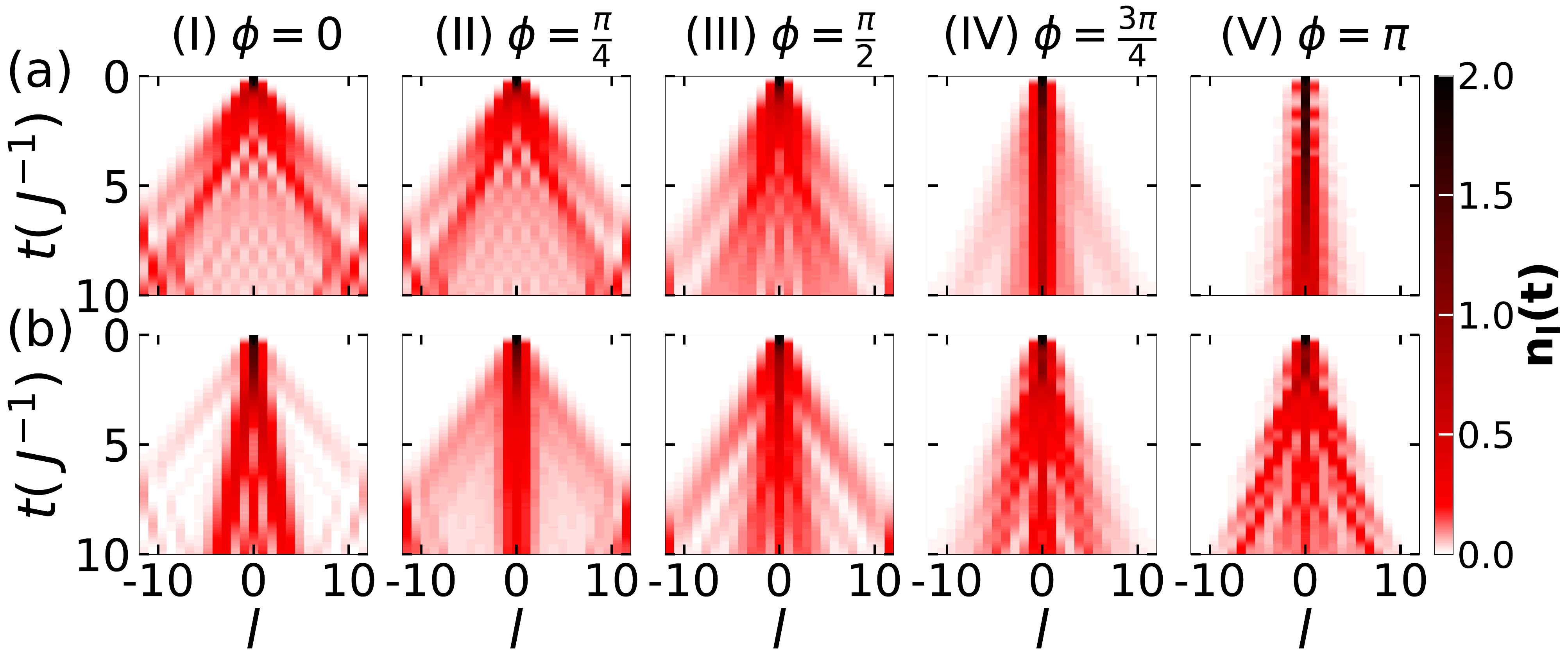}
	\caption{$n_l(t)$ at each rung plotted as a function of time $t$(in units of $J^{-1}$) with the initial state $|\Psi(0)\rangle = b_{0, A}^\dagger b_{0, B}^\dagger |vac\rangle$. Panel (a) and (b) show the density evolution for interaction strength $U=4$ and $U=20$, respectively for different values of $\phi$. }
	\label{fig:tp QW}
\end{figure}

We study the QW following the standard protocol of unitary time evolution as 
\begin{equation}
\left|\Psi(t) \right\rangle = U(t) \left|\Psi(0)\right\rangle,
\label{eqn:qw}
\end{equation}
where $U(t) = e^{-iHt/\hbar}$ and $\left|\Psi(0)\right\rangle$ is an initial state at time $t=0$. The analysis is done by obtaining the exact numerical solution of 
Eq.~\ref{eqn:qw}.
In our calculations, we use the hopping strengths $J=1$ and $K=3$ as considered in the experiment of Ref.~\cite{Tai_2017} and study the QW by varying $U$ and $\phi$. Here the choice of $J=1$ sets the energy scale in our studies. The calculations are done with a ladder of $L=25$ rungs which is a system of $50$ lattice sites.

\begin{figure}[t]
	\centering	\includegraphics[width=0.95\columnwidth]{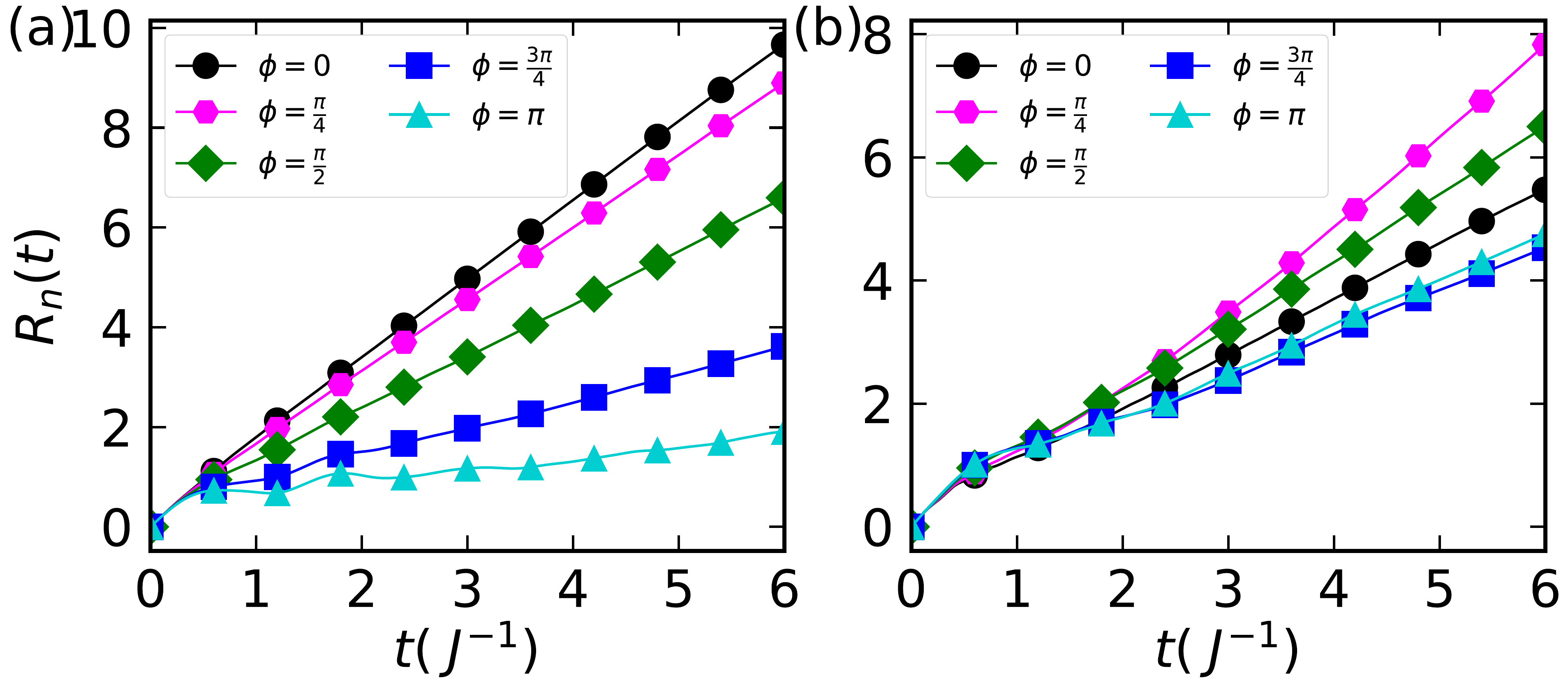}
	\caption{ (a-b) Show the radius of the expansion plot as a function of time $t$(in units of $J^{-1}$) for interaction strength $U=4$ and $U=20$.}
	\label{fig:Band_corr}
\end{figure}

\section{Results}
In this section we present our main results. For the QW, we consider the initial state $|\psi(0)\rangle=b_{0,A}^\dagger b_{0,B}^\dagger|vac\rangle$,
which corresponds to a state with two-particles created at the central rung of the ladder with one particle at each site of the rung. It is to be noted that the dynamics of this state under the influence of uniform flux ($\phi$) and fixed onsite interaction strength ($U$) and fixed rung-to-leg hopping ratio ($K/J$) has been discussed in Ref.~\cite{Tai_2017}, where the focus was primarily on the chiral motion of the particles.
In this work we systematically investigate the combined role of flux and interaction both in weak to strong regimes on the two-particle dynamics where we focus on both the radial as well as the chiral dynamics.

\begin{figure}[t]
	\centering
	\includegraphics[width=1.0\columnwidth]{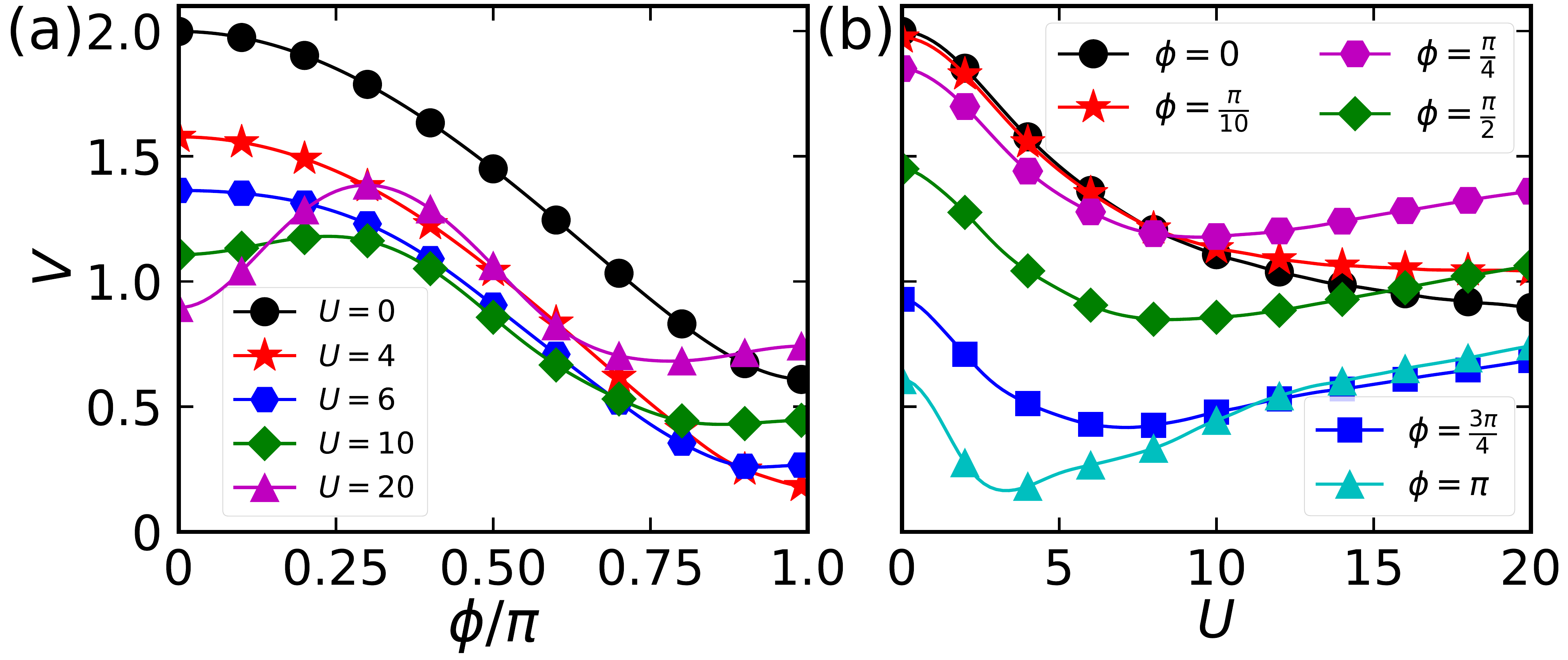}
	\caption{The figure (a) and (b) show the radial velocity $V$ of the wavefunction as a function of $\phi/\pi$ and $U$ in presence of uniform flux for different values of $U$ and $\phi$. For the calculation of $V$ we consider $t_f=5J^{-1}$ and $t_i=2J^{-1}$.}
	\label{fig:velocity}
\end{figure}

First of all, to understand the radial dynamics, we examine the density evolution of the particles. In Fig.~\ref{fig:tp QW}(a) and (b), we show the time evolution of the total onsite density in a rung i.e. $n_l(t) =\langle n_{l,A}(t)\rangle + \langle n_{l,B} (t)\rangle$ for $U=4$ and $U=20$ respectively and for each case we have considered $\phi=0,~\pi/4,~\pi/2,~3\pi/4$ and $\pi$. In the presence of weak interaction ($U=4$) we find the spreading of density is suppressed with an increase in flux strengths as shown in Fig.~\ref{fig:tp QW}(a).
However, when the interaction is strong, we see a non-monotonous behaviour in the spreading of the densities with an increase in $\phi$ as shown in Fig.~\ref{fig:tp QW}(b) for $U=20$. In order to clearly understand this behaviour, we compute the radial velocity, $V = \frac{R_n(t_f)-R_n(t_i)}{t_f-t_i}$ from the slope of the linear region in the radius of expansion defined as $R_n(t)=[\sum_{l}(l-l_0)^2\langle n_l(t)\rangle]^{1/2}$ (where $l_0$ is the index of the central rung of the ladder) for different $U$ and plot them as a function of $\phi/\pi$ in Fig.~\ref{fig:velocity}(a).
This figure depicts two important information: 

(i)For small values of $U$ i.e. $U=0$ ( black circle), $U=4$ (red stars) and $U=6$ (blue hexagons), the radial velocity $V$ decreases smoothly as a function of $\phi/\pi$. However, for stronger interactions i.e. $U=10$ (green diamonds) and $U=20$ (magenta triangles),  $V$ first increases then decreases and increases again with increase in $\phi$ indicating a re-entrant behaviour in the radial dynamics.
(ii)For fixed values of $\phi$,  $V$ decreases monotonously with $U$ when $\phi$ is small. After a certain value of $\phi$, it first decreases and then increases with increase in $U$ indicating another re-entrant behaviour which is depicted in Fig.~\ref{fig:velocity}(b).

From the above results it is clear that the dynamics as a function of $\phi$ when $U=0$ is similar to the single particle case. In this case, all the available states are scattering states that contribute to a fast dynamics when $\phi=0$. Increase in $\phi$ increases the flatness of the band or in other words decreases the band width resulting in the slow dynamics\cite{Tokuno_2014,Paredes_2014,oktel2015,Rashi_2017}. However, the re-entrant feature in $V$ in the strong interaction regime can be understood by first understanding the decrease in $V$ as a function of $U$ in the absence of flux. As depicted in Fig.~\ref{fig:velocity}(a), $V$ smoothly decreases as a function of $U$ when $\phi=0$ (also see Fig.~\ref{fig:velocity}(b)). This slowing down of dynamics happens solely due to the effect of interaction and this can be attributed to the formation of some kind of bound state in the system. In order to gain insights about such possibilities, we plot the two particle energy band diagram in Fig.~\ref{fig:Band_corr} (upper panel). To determine the band structure, we consider the periodic boundary condition and perform an exact diagonalization (ED) calculation in the momentum space~\cite{wen2021, ke2017}.

For $U=1$, two well defined bands of scattering states exist as shown in Fig.~\ref{fig:Band_corr}(a). As $U$ increases, isolated bands corresponding to the bound states start to appear above each scattering state bands which are shown in Fig.~\ref{fig:Band_corr}(b) and  Fig.~\ref{fig:Band_corr}(c) for $U=4$ and $U=20$ respectively. These isolated bands shift towards higher energies with increase in $U$  and become more and more flat (compare Fig.~\ref{fig:Band_corr}(b) and Fig.~\ref{fig:Band_corr}(c)). The flatness of the bound state bands with increase in $U$ causes the slowing down of dynamics or decrease in the radial velocity $V$. 

To identify this bound state, we compute the time-evolved two-particle correlation function defined as $\Gamma_{i, j}=\langle b_i^\dagger b_j^\dagger b_j b_i\rangle$, here $b_i(b_i^\dagger)$ is the particle annihilation(creation) operator and $i, ~j$ are the site index of the ladder. For this calculation, the indexing starts from the left most site of leg-B of the ladder such that even (odd) indices are on leg-B (leg-A). 
In Fig.~\ref{fig:Band_corr}(d-f), we plot $\Gamma_{i,j}$ for $U=1,~4,~20$ respectively at $t=4J^{-1}$ of the time evolution of the two-particle initial state $|\Psi(0)\rangle$. This clearly shows that for $U=1$, the two particles perform independent particle quantum walk exhibited by symmetric distribution of the correlation matrix elements as shown in Fig.~\ref{fig:Band_corr}(d)~\cite{Greiner_walk,lahini2012qw,mondalwalk}. However, with finite and small interaction ($U=4$), the correlation matrix elements $\Gamma_{i, j}=\langle b_i^\dagger b_{j}^\dagger b_{j} b_i\rangle$  become dominant when $i$, and $j$ sites are on same rung which can be seen as the dark spots in Fig.~\ref{fig:Band_corr}(e).  
This feature indicates the dynamics of a bound state formed between the particles on the two sites of the rung which we call the rung-pair state. At this moderate interaction strength, we also see signatures of finite off diagonal matrix elements which are the contributions from the single particle dynamics that arise due to the presence of the nearby scattering state band (compare Fig.~\ref{fig:Band_corr} (c)). For very strong values of interaction ($U=20$), we see a clear dominance of the rung-pair state in the dynamics as shown in Fig.~\ref{fig:Band_corr}(f). Formation of this rung-pair state is the reason behind the slow dynamics as a function of interaction. 
Note that the other bound state band at higher energy does not have significant contribution in the dynamics. 
\begin{figure}[t]
	\centering	\includegraphics[width=1.0\columnwidth]{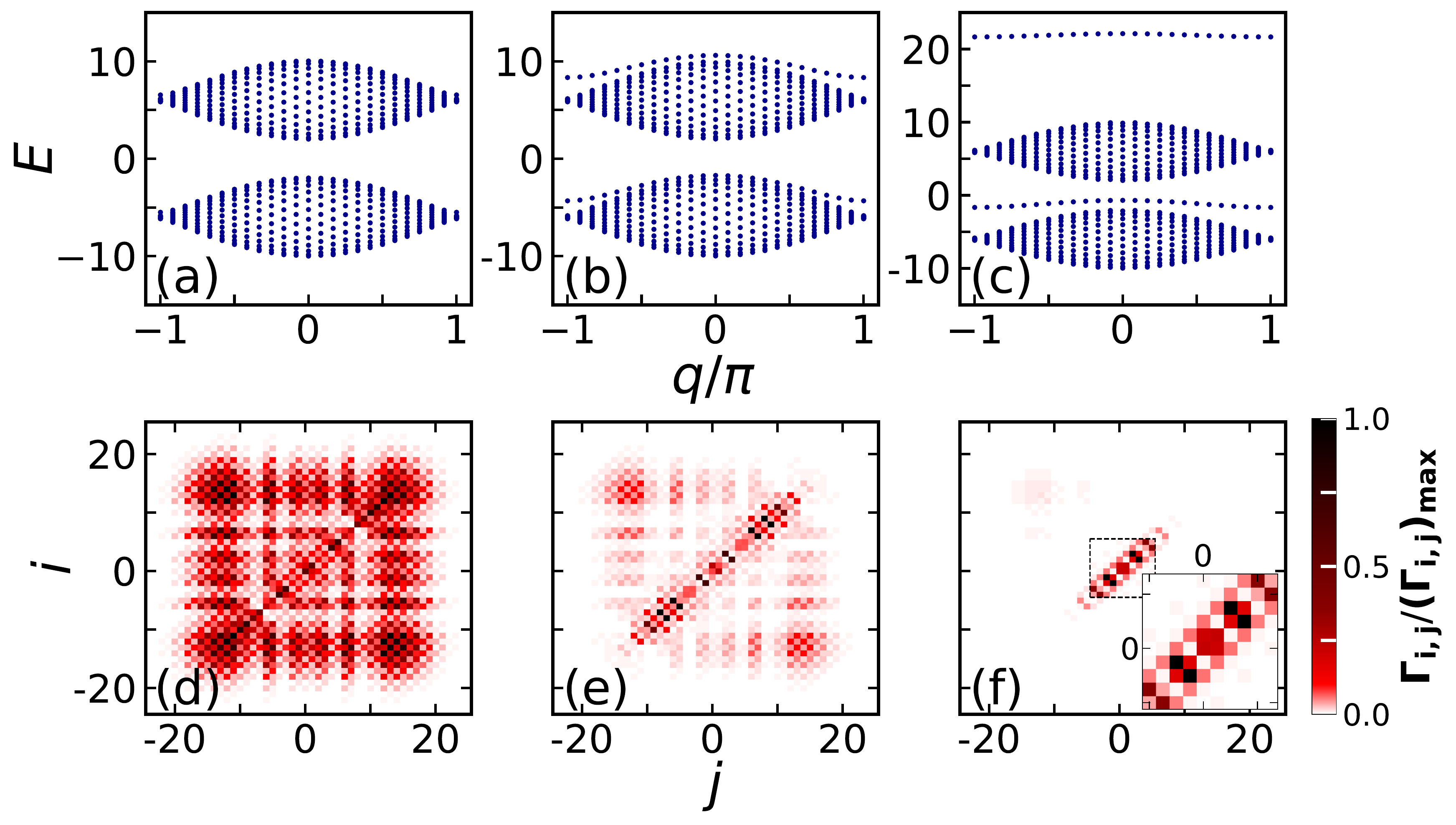}
	\caption{ (a-c) Show the energy band diagram of the two particle system for $U=1$, $U=4$ and $U=20$ respectively in the absence of $\phi$ with $q$ as the quasi-momentum. (d-f) Show the two particle correlation $\Gamma_{i,j}$ at time $t=4J^{-1}$ for the same parameters used respectively in (a-c). Here $i$ and $j$ are the site indices.}
	\label{fig:Band_corr}
\end{figure}

\begin{figure*}[t]
	\centering
	\includegraphics[width=2.0\columnwidth]{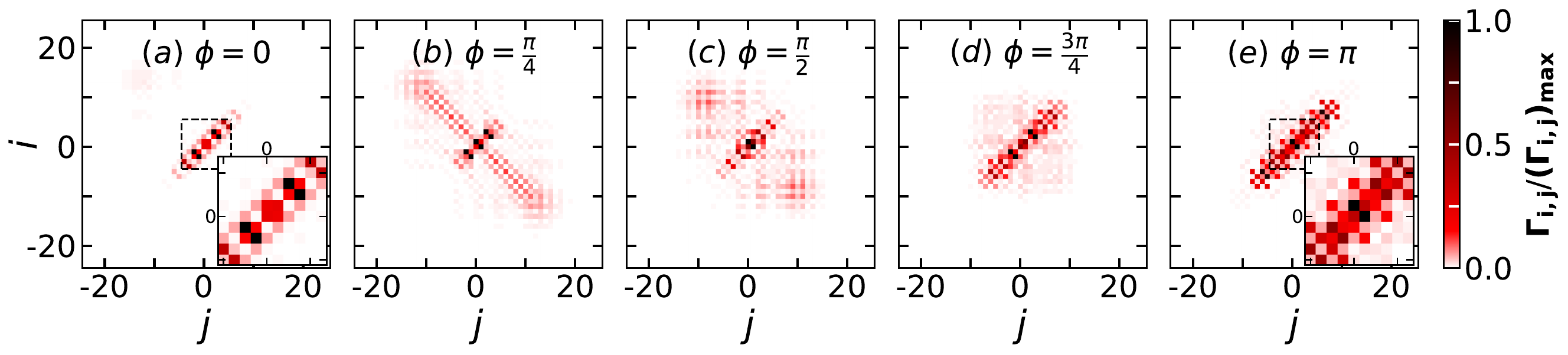}
	\caption{Figure shows the two-particle correlation $\Gamma_{i,j}$ for different values of $\phi$ and $U=20$ plotted at time $4J^{-1}$ of the evolution. In each case we have normalized the correlation as $\Gamma_{i,j}/(\Gamma_{i,j})_{max}$ for clarity. Here $i$ and $j$ are the site indices.}
	\label{fig:corr_U_20}
\end{figure*}

\begin{figure}[!b]
	\centering
	\includegraphics[width=1.0\columnwidth]{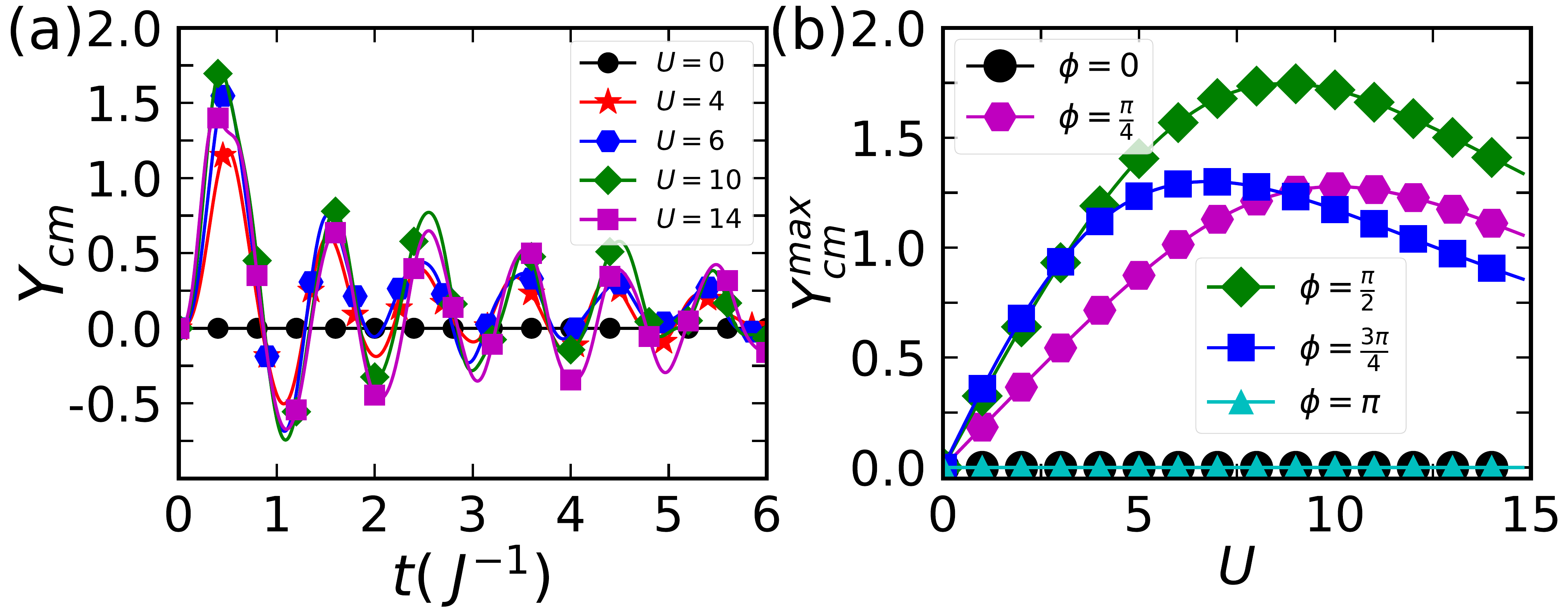}
	\caption{(a) The shearing, $Y_{CM}$ is plotted as a function of time $t$(in units of $J^{-1}$) for different values of $U$ at $\phi = \frac{\pi}{2}$. (b) The maximum of the shearing, $Y_{cm}^{max}$ is plotted as function of $U$ for different $\phi$ values.}
	\label{fig:TP Ycm}
\end{figure}
From the insights obtained from the above analysis, we can now understand the re-entrant behaviour in $V$ as a function of $\phi$ in the regime of strong interaction as already depicted in Fig.~\ref{fig:velocity}(a). In this regime, when a small flux is introduced into the system, there is a finite probability of rung-pair breaking due to the onset of the chiral dynamics~\cite{Tai_2017} and therefore both rung-pair and independent particle QW is expected. This can be seen as the finite elements in the time-evolved correlation matrix $\Gamma_{i,j}$ away from the main diagonal along with the contribution from the rung-pair dynamics as shown in Fig.~\ref{fig:corr_U_20}(b) for $U=20$ and $\phi=\pi/4$ (see Fig.~\ref{fig:corr_U_20}(a) for comparison). As a result of this contribution from the independent particle QW, $V$ increases as a function of $\phi$ in the weak flux regime (see Fig.~\ref{fig:velocity}(a) for $U=20$). However, with further increase in $\phi$ the independent particle QW gets affected due to the band-flattening~\cite{Tokuno_2014,Paredes_2014,oktel2015,Rashi_2017} and $V$ decreases - a situation similar to the case of $U=0$ as depicted in Fig.~\ref{fig:velocity}(a). Note that at $\phi/\pi\sim 0.75$, $V$ reaches a minimum where it is smaller than the value at $\phi=0$. This reduction in velocity can be inferred from the behaviour of $\Gamma_{i,j}$ shown in Fig.~\ref{fig:corr_U_20}(c) and Fig.~\ref{fig:corr_U_20}(d) for $\phi=\pi/2$ and $\phi=3\pi/4$ respectively where the elements far from the diagonal become smaller and smaller with $\phi$.
Interestingly, further increase in $\phi$ favours the formation of the rung-pair again as can be seen from Fig.~\ref{fig:corr_U_20}(e) for $\phi=\pi$. This time the contributions from the free particle dynamics is suppressed and we see an increase in the nearest neighbour elements along the diagonal of the correlation matrix indicating a strong rung-pair dynamics again. This feature reveals that the higher values of $\phi$ favour the formation of rung-pair state in the regime of strong interaction and strong rung-to-leg hopping ratio. As a result of this $V$ increases again leading to the re-entrant behaviour ( Fig.~\ref{fig:velocity}(a) for $U=20$). The increased probability of the rung-pair state formation can be attributed to the formation of closely spaced vortices in the ladder~\cite{dhar2012,Sebastian2015} in the limit of strong flux due to which the probability of finding a boson pair delocalized on the rungs increases.
Note that due to the closely spaced vortices in the system, we expect finite correlations along the rungs and also along the diagonals of the plaquettes which are shown in Fig.~\ref{fig:corr_U_20}(e) for $\phi=\pi$ and this feature is absent in Fig.~\ref{fig:corr_U_20}(a) for $\phi=0$. It is to be noted that this re-entrant dynamics in the strong interaction limit is possible only when $K\gtrsim 2J$ at which the band corresponding to the rung-pair state starts to separate from the scattering bands (not shown) - a condition favourable for the rung-pair formation.

We now proceed to investigate the chiral nature introduced by $\phi$ and $U$ in the two-particle dynamics~\cite{Tai_2017}. 
To  quantify the chirality, we calculate the shearing along the $y$-direction of the ladder $Y_{CM} = N_{R} -N_{L}$~\cite{Tai_2017, Hui2017} 
where, $N_{R} = \frac{\sum\limits_{l>0} (n_{l,B}-n_{l,A})}{\sum\limits_{l>0} (n_{l,B}+n_{l,A})} \;\;\text{and}\;\;
    N_{L} = \frac{\sum\limits_{l<0} (n_{l,B}-n_{l,A})}{\sum\limits_{l<0} (n_{l,B}+n_{l,A})}$.
We plot $Y_{CM}$ as a function of time for different values of $U$ (i.e. $U=0,~4,~6,~10,~14$) in  Fig.~\ref{fig:TP Ycm}(a) for a fixed value of $\phi=\pi/2$. The figure shows that $Y_{CM}=0$ when $U=0$ (black circle) which indicates the absence of chirality in the system~\cite{Tai_2017}. However, for $U \neq 0$, a finite oscillation in $Y_{CM}$ starts to appear due to the onset of chirality in the dynamics. Interestingly, the amplitude of oscillation in $Y_{CM}$ gradually increases and then decreases with increase in $U$ in the short time dynamics. To quantify this we plot the maximum value of $Y_{CM}$ ($Y_{CM}^{max}$) after a short time evolution as a function of $U$ for different values of $\phi$ in Fig.~\ref{fig:TP Ycm}(b). The increase and then decrease in the value of $Y_{CM}^{max}$ with interaction $U$ for $\phi=\pi/4$ (magenta hexagons), $\pi/2$ (green diamonds) and $3\pi/4$ (blue squares) is an indication of a re-entrant chiral dynamics. Note that the chirality is absent for $\phi=0$ (black circles) and $\phi=\pi$ (cyan triangles).
\begin{figure}[b]
	\centering        
 \includegraphics[width=1.\columnwidth]{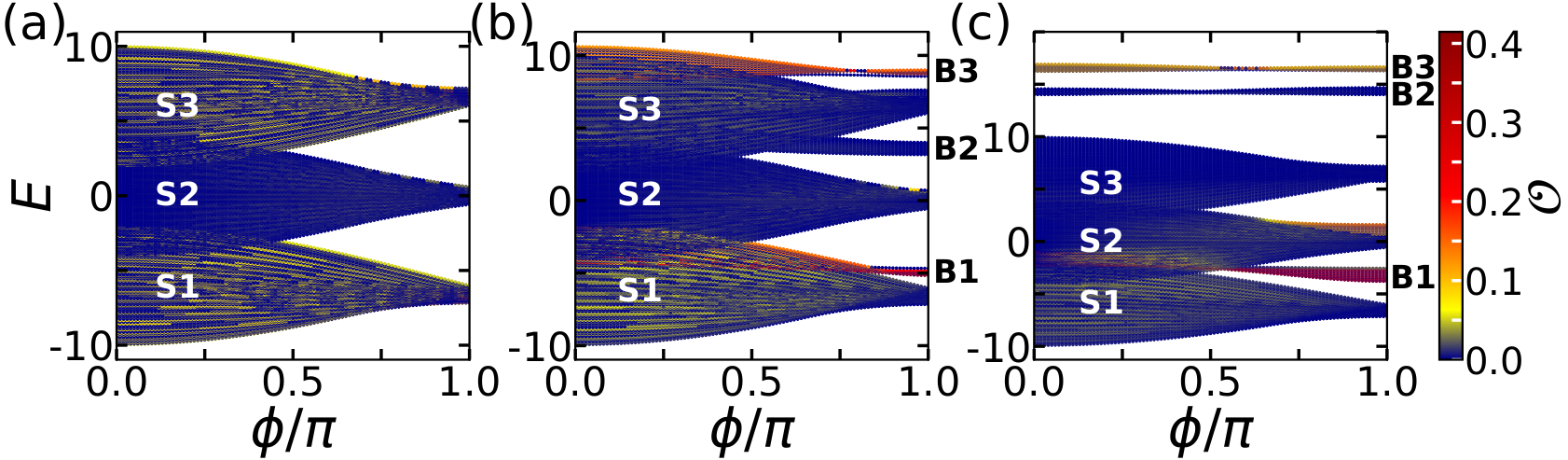}
	\caption{Two particle energy spectrum of the ladder as a function of $\phi$. The color scale represents the overlap $O$ of the initial state $|\Psi(0)\rangle= b_{0,A}^\dagger b_{0,B}^\dagger |vac\rangle$ with all the energy eigenstates of the system. Figures (a), (b) and (c) correspond to the uniform flux with $U=0,4,$ and $14$, respectively.}
	\label{fig:Plot-set1}
\end{figure}

To understand this behaviour in chirality we analyze the energy eigenstates of the system. As already discussed in Ref.~\cite{Tai_2017}, for the choice of the initial state $|\Psi(0)\rangle$, when $U=0$, the two bosons do not exhibit chiral motion in their QW due to the symmetric population of the chiral bands. In contrast, finite interaction $U$ breaks this symmetry and the chiral motion sets in. This can be understood by calculating the overlap of the initial state $|\Psi(0)\rangle$ with all the two particle energy eigenstates ($\{|\chi_{i}\rangle\}$) of the Hamiltonian defined as
    $\mathcal{O} = |\langle \chi_i | \Psi(0)\rangle|.$

We plot $\mathcal{O}$ as a function of $\phi/\pi$ in Fig.~\ref{fig:Plot-set1} (a-c) for $U=0,~4,~14$ respectively. 
In each cases we obtain three scattering bands namely, the S1, S2 and S3 bands. 
As already known, the chiral dynamics is mainly due to the contribution from the S1 and S3 scattering bands, whereas the S2 scattering band does not contribute to chirality~\cite{Tai_2017}. It can be seen that for $U=0$ (Fig.~\ref{fig:Plot-set1} (a)) and $U=4$ (Fig.~\ref{fig:Plot-set1} (b)), the overlap $\mathcal{O}$ with the S1 and S3 bands are significant. While the overlap with the S1 and S3 bands for $U=0$ is symmetric, for $U=4$ an asymmetric overlap  is seen which sets in the chirality in the system (compare Fig.~\ref{fig:TP Ycm}). This asymmetry in $\mathcal{O}$ increases  as the interaction increases further resulting in the increase in chirality. However, further increase in $U$, the overlap $\mathcal{O}$ with the S2 band starts to become finite as can be seen from Fig.~\ref{fig:Plot-set1}(c) for $U=14$. This reduces the chirality in the system and as a result a re-entrant feature in chiral dynamics appears. Note that three bound state bands (B1, B2 and B3) also appear for finite $U$ but they don't contribute to the chiral dynamics.

\section{Conclusion}
In this work, we have investigated the QW of two interacting bosons on a two-leg ladder in the presence of uniform flux. 
Starting from an initial state having one particle on each site of the central rung of the ladder, we have shown that in the regime of dominant rung-hopping and strong interaction an interesting re-entrant dynamics occurs which exhibits a slow spreading wavepacket becomes fast then slow and fast again as a function of the flux strength. However, in the limit of weak interaction the dynamics is monotonous.  Moreover, we have obtained a re-entrant feature in the two particle chiral dynamics as a function of interaction where we have found that the center of mass shearing first increases and then decreases for different values of flux strengths. 
Note that although the re-entrant dynamics in the interacting QW has been predicted in systems with two competing interactions by tuning the interaction strength~\cite{mondalwalk,giri2022nontrivial,Longhi_BO_2012,Lin_BO_2014}, in our work we have predicted a re-entrant dynamics mediated by an external gauge field and not by any interaction for the first time. 

Our study reveals in the context of the quench dynamics in a system of two-leg flux ladder which is one the most discussed model in recent years which has been studied extensively both theoretically and experimentally. The QW of interacting bosons on a two-leg flux ladder has also been analysed in a recent experiment using cold atoms in optical lattices~\cite{Tai_2017}. Therefore, our findings can be observed a an existing experimental setup and also may open of new directions where investigations can be made to understand this re-entrant dynamics with different types of initial conditions as well as perturbations such as interaction and disorder.  

{\em Acknowledgements.-}
We thank Subroto Mukerjee and Suman Mondal for useful discussions. T.M. acknowledges support from Science and Engineering Research Board (SERB), Govt. of India, through project No. MTR/2022/000382 and STR/2022/000023.  

\bibliography{references}
\end{document}